\documentclass[physrev,twocolumn,floatfix,superscriptaddress,longbibliography,floatfix,hyperlinks]{revtex4-1}

\usepackage{graphicx}
\usepackage{float}
\usepackage{times}
\usepackage[colorlinks=true,citecolor=blue]{hyperref}
\usepackage{soul}
\usepackage{ulem}
\usepackage{amsmath}
\usepackage{amssymb}
\usepackage{mathrsfs}
\usepackage[usenames,dvipsnames]{color}

\begin{document}

\title{Designer Flat Bands in Quasi-One-Dimensional Atomic Lattices}

\author{Md Nurul Huda}
\affiliation{Department of Applied Physics, Aalto University, FI-00076 Aalto, Finland.}

\author{Shawulienu Kezilebieke}
\affiliation{Department of Applied Physics, Aalto University, FI-00076 Aalto, Finland.}

\author{Peter Liljeroth}
\email{Email: peter.liljeroth@aalto.fi}
\affiliation{Department of Applied Physics, Aalto University, FI-00076 Aalto, Finland.}

\date{\today}
\begin{abstract}


Certain lattices with specific geometries have one or more spectral bands that are strictly flat, i.e.~the electron energy is independent of the momentum. This can occur robustly irrespective of the specific couplings between the lattices sites due to the lattice symmetry, or it can result from fine-tuned couplings between the lattice sites. While the theoretical picture behind flat electronic bands is well-developed, experimental realization of these lattices has proven challenging. Utilizing scanning tunnelling microscopy (STM) and spectroscopy (STS), we manipulate individual vacancies in a chlorine monolayer on Cu(100) to construct various atomically precise 1D lattices with engineered flat bands. We realize experimentally both gapped and gapless flat band systems with single or multiple flat bands. We also demonstrate tuneability of the energy of the flat bands and how they can be switched “on” and “off” by breaking and restoring the symmetry of the lattice geometry. The experimental findings are corroborated by tight-binding calculations. Our results constitute the first experimental realizations of engineered flat bands in a 1D solid-state system and pave the way towards the construction of e.g.~topological flat band systems and experimental tests of flat-band-assisted superconductivity in a fully controlled system.

\end{abstract}

\maketitle 
\section{Introduction}
There has been a surge of interest in systems exhibiting flat electronic bands, i.e. structures where one (or more) of the bands are completely dispersionless throughout the Brillouin zone \cite{Leykam2018_AdvPhysX,Su2010,Bistritzer2011_TBG,Jarillo-Herrero2018_TBG_SC}. Flat band (FB) systems have existed as theoretical proposals already for a long time \cite{Sutherland1986_PRB,lieb1989theorems,Mielke_1991a,Mielke_1991b,Tasaki1992_PRL,Mielke1999_PRL,Vidal2000_PRL,Tasaki2008_review,Heine2020_ChemSocRev}, but only now systems exhibiting such response can be fabricated and studied experimentally. The quenched kinetic energy in a flat band (FB) causes localization, the wave group velocity goes to zero. The zero kinetic energy also means that any other energy scale can be the dominant one: flat bands are not stable against perturbations and even weak interactions can induce the formation of broken symmetry ground states, such as ferromagnetism \cite{Mielke1999_PRL,lieb1989theorems,Costa2016_PRB,Derzhko2010_PRB,Tasaki2008_review}, Wigner crystals \cite{Wu2007_PRL}, superconductivity \cite{Kopnin2011,Peotta2015_NatCommun,julku2016,lothman2017_PRB}, or fractional quantum Hall, quantum anomalous Hall, and fractional Chern insulator states \cite{Sun2011_PRL,Tang2011_PRL,Zhao2012_PRB,Jaworowski2015_PRB}. In addition, the extremely high density of states of the flat band can enhance the transition temperature to, e.g., the superconducting phase \cite{Kopnin2011}. In addition to the theoretical proposals, flat bands have also been realized experimentally in optical systems and ultracold atomic gases \cite{Gersen2005_PRL,Jacqmin2014_PRL,mukherjee2015,Vicencio2015_PRL,taie2015,Nguyen2018_PRL}, in electronic systems arising in real materials (notably twisted bilayer graphene) \cite{Jarillo-Herrero2018_TBG_SC,Li2018_SciAdv,Marchenko2018_SciAdv,po2018origin}, and recently, in artificial materials fabricated through atom manipulation with the scanning tunneling microscope (STM) \cite{Drost:NatPhys2017,Slot:2017NatPhys,Slot2019_PRX,Gardenier2020_ACSNano}. The artificial systems are extremely interesting, as they allow complete tuneability of the lattice symmetry and intentional introduction of additional effects, such as defects or disorder. The systems that have been demonstrated are based on the two-dimensional Lieb lattice \cite{lieb1989theorems,Drost:NatPhys2017,Slot:2017NatPhys}, where the flat band results from the lattice symmetry. However, it is only completely dispersionless if the next-nearest neighbour (NNN) hoppings are negligible, which is unlikely in a condensed-matter system \cite{Drost:NatPhys2017,Slot:2017NatPhys}.

\begin{figure*}[t!]
    \centering
    \includegraphics{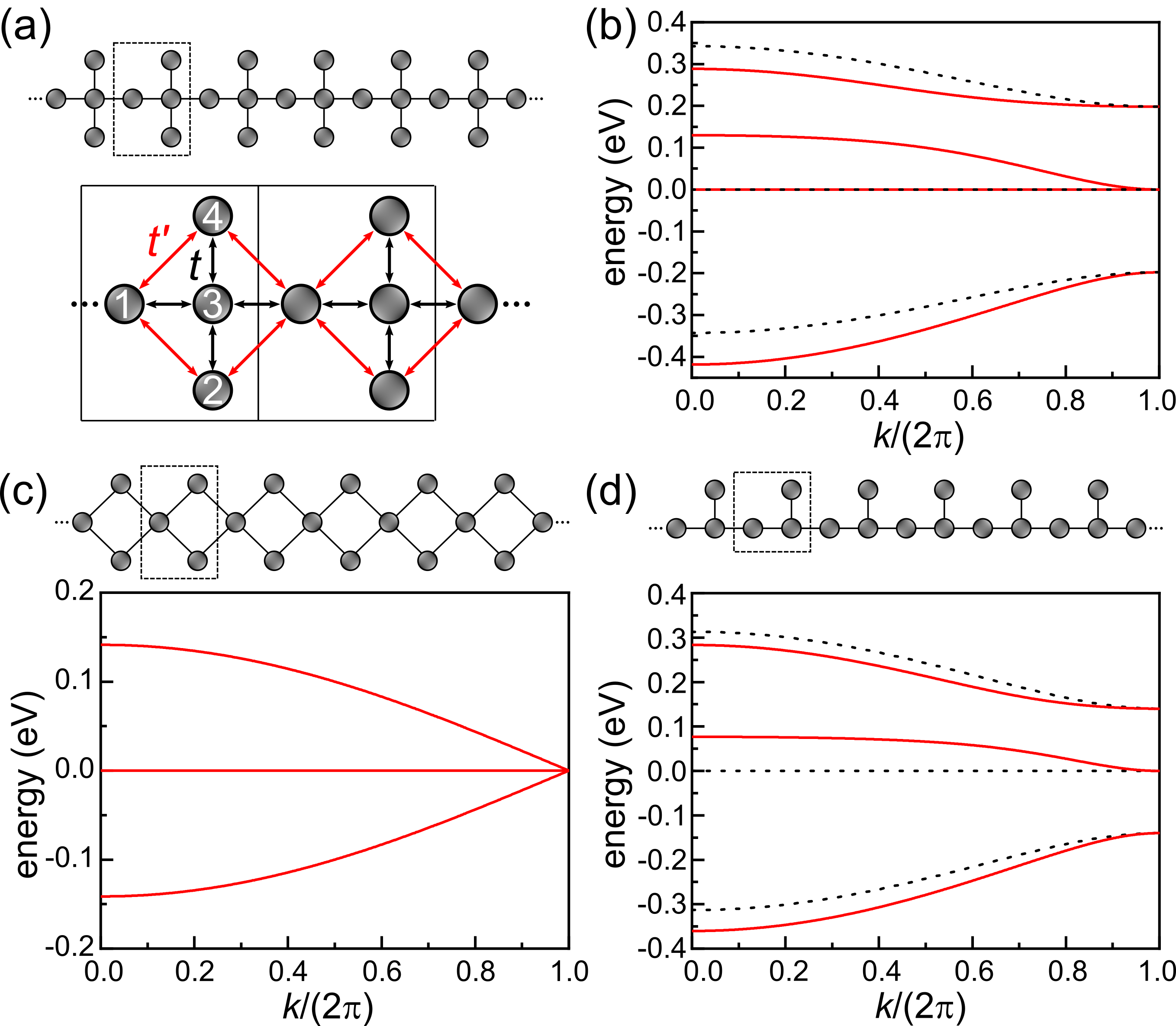}    
    \caption{1D flat band lattice structures. (a) Schematic of 1D cross lattice structure and a zoom-in showing the nearest-neighbour ($t$) and NNN hoppings ($t'$). (b) Calculated energy spectrum of the 1D cross lattice. (c) Schematic of 1D diamond lattice structure (top) and calculated energy spectrum of the lattice (bottom). (d) Schematic (top) and calculated energy spectrum (bottom) of the stub lattice structure. The energies are given w.r.t.~the on-site energy. The dotted and solid lines represent the band structures for zero and non-zero NNN hoppings.}
    
    \label{fig:1}
\end{figure*}

In addition to the iconic 2D Lieb lattice, there are several text-book examples of 1D systems that exhibit flat bands: diamond, cross and stub chains (see Fig.~\ref{fig:1}) \cite{Miyahara2005_JPSJap,Hyrkas2013_PRA,Morales-inostroza2016}. These are all examples of systems where the flat bands stem from the local lattice symmetry rather than requiring fine-tuning of the various couplings between the lattice sites. This ensures the existence of one (or a few) completely dispersionless bands in the spectrum and relies on the existence of compact localized eigenstates due to destructive interference, enabled by the local symmetries of the network \cite{Aoki1996_PRB,Khomeriki2016_PRL}. These compact localized eigenstates can be engineered, i.e. it is possible to design systems exhibiting flat bands in 1D, 2D, or 3D \cite{deng2003,Miyahara2005_JPSJap,Apaja2010_PRA,Morales-inostroza2016,Leykam2018_AdvPhysX,Rhim2019_PRB,Kuno_2020,Lima2020_PRB}. While it is not a priori clear if the NNN interactions cause the flat band to become dispersive, it is possible to incorporate this into the design principles in 1D to ensure robustness of the flat bands \cite{Morales-inostroza2016}. 

Currently, there is theoretical progress towards incorporating and studying additional effects, such as topological flat band systems, effects of interactions and disorder or fractal-like geometries \cite{Sun2011_PRL,Tang2011_PRL,Zhao2012_PRB,Jaworowski2015_PRB,Bodyfelt2014_PRL,Bercioux2017_AnnPhys,Pal2018_PRB,Platero2019_FB}. However, there is a need for a flexible experimental platform where the various design principles can be tested, verified and eventually exploited. While the platforms based on photonic and optical lattices are tuneable and extremely well-controlled \cite{mukherjee2015,taie2015,Jacqmin2014_PRL,Vicencio2015_PRL,Gersen2005_PRL,Nguyen2018_PRL}, it would be of interest to be able to demonstrate these effects in an electronic system that can be precisely controlled. Artificial lattices based on atom manipulation by the tip of an STM have emerged as highly tuneable system for studying the effect of the lattice symmetry and couplings on the resulting band structure \cite{Gomes2012,Drost:NatPhys2017,Slot:2017NatPhys,Slot2018,Slot2019_PRX,freeney2019edge,Yan2019_NJP,Yan2019_AdvPhysX,Gardenier2020_ACSNano}. Here, we will use the chlorine vacancy system on Cu(100) \cite{Kalff:NatNano2016,Girovsky:SciPost2017,Drost:NatPhys2017,Huda2020_npj} for constructing various 1D lattices with engineered flat bands. We will focus experimentally on the effects of the NNN hoppings to test the robustness of the engineered bands, how to engineer both gapped and metallic flat-band systems, and to demonstrate how by breaking the symmetry of the chain geometry, we can control the dispersion of the flat bands. These result constitute the first experimental realizations of engineered flat bands in a 1D solid-state system and pave the way towards the construction of e.g.~topological flat band systems and experimental tests of flat-band-assisted superconductivity in a fully controlled system \cite{Imada200PhysRevLett,Kobayashi2016PhysRevB,Bercioux2017_AnnPhys,yankowitz2019_Sci}.

\section{Results and Discussion}
\subsection{Model results on flat bands in one-dimensional chains}
In order to illustrate the ideas on how to construct flat bands, consider the 1D cross lattice shown in Fig.~\ref{fig:1}(a). The calculated band structures of this system using a nearest- and NNN tight-binding (TB) models are shown in Fig.~\ref{fig:1}(b) (energies given w.r.t.~the on-site energy) with the hoppings that correspond to the experimentally determined values (see below). Only including the NN hoppings (black arrows in Fig.~\ref{fig:1}(a)) results in a band structure with a doubly degenerate flat band at the mid-gap energy between two dispersive bands (dashed line in Fig.~\ref{fig:1}(b)). The flat band in this structure appears as a solution where the single-particle wave function is zero at the connecting sites of the lattice, making it impossible to have transport through the lattice. Specifically, this corresponds to having zero wavefunction intensity on site 3 in the unit cell of the cross lattice. Turning on NNN hoppings can cause the flat band to acquire dispersion; in the case of the cross chain, one of the flat bands becomes dispersive while the other one remains flat. Looking at the wavefunctions corresponding to the flat band, they now have zero intensity on sites 1 and 3. When the NNN hoppings (red arrows in Fig.~\ref{fig:1}(a)) are included, it is necessary to have zero intensity on these two sites to completely block transport through the unit cell.

Other typical 1D flat-band systems include the diamond (Fig.~\ref{fig:1}(c)) and stub lattices (Fig.~\ref{fig:1}(d)). If we include only the nearest-neighbour hoppings, all of the above systems host flat bands that are pinned to the on-site energy. These lattices are typical examples of systems with flat bands: we can recognize that they all are bipartite lattices with sites with different connectivities \cite{Leykam2018_AdvPhysX}. For example in the case of the diamond lattice, it has ``rim'' sites which only have two nearest neighbours and ``hub'' sites that have four nearest neighbours. In addition to the flat band, the diamond lattice has Dirac-like bands that touch at the edge of the 1D Brillouin zone. The cross and the stub chain have gapped spectra with the flat band at the mid gap energy. Turning on NNN hoppings can cause the flat band to acquire dispersion. This is what happens in the stub chain (Fig.~\ref{fig:1}(d)), while in the cross chain, as discussed above, only one of the flat bands becomes dispersive (Fig.~\ref{fig:1}(b)). In the case of the diamond chain, NNN hoppings between top and bottom sites in the unit cell do not give dispersion to the flat band. If we include NNN hoppings between adjacent unit cells, the flat band becomes dispersive (not shown). However, in our experimental realization (described below), the NNN hoppings are much weaker in the diamond chain (due to the distance between the sites). These TB calculations demonstrate that it should be possible to create engineered 1D structures with flat bands, even as nearest-neighbour interactions cannot be ignored in experimental structures.

\begin{figure*}[t!]
	\centering
	\includegraphics{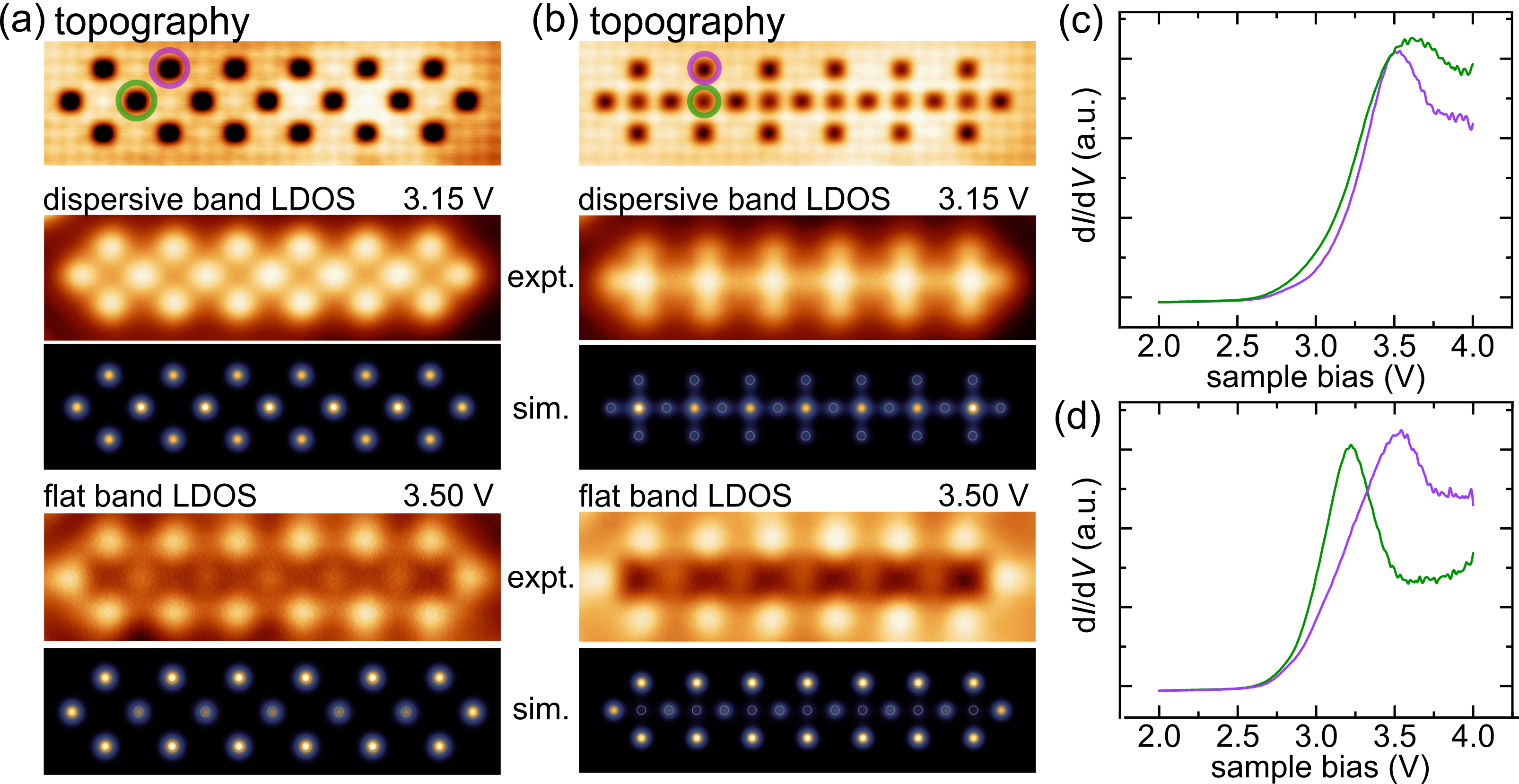}  
	\caption{Experimental realization of the flat band lattices. (a,b) The topography (top) and the dispersive and flat band LDOS maps of the 1D diamond (a) and cross chains (b). The simulated LDOS maps at the corresponding energies are shown for comparison. (c,d) d$I$/d$V$ spectra taken on the sites with maximum flat band (purple) and dispersive band (green) intensity of 1D diamond (c) and cross chains (d). The location of the spectra are indicated in the STM topography images in panels a and b.}
	\label{fig:2}
\end{figure*}

\subsection{Experimental realization of diamond, cross and stub chains}
In order to realize the flat bands experimentally, we fabricated 1D lattice structures by atomic manipulation in chlorine vacancy system introduced in Refs.~\cite{Kalff:NatNano2016,Drost:NatPhys2017,Girovsky:SciPost2017,Huda2020_npj}. The sample preparation and the experimental methods are described in detail in Appendix~\ref{Expt}. Fig.~\ref{fig:2} shows the  experimental realization of 1D flat bands in diamond and cross lattice structures. The LDOS of the sample is measured by d$I$/d$V$ signal in constant-height mode. Along with the STM topography, Fig.~\ref{fig:2}(a) shows the experimental and simulated LDOS map of the dispersive and flat bands of the diamond chain. LDOS can be simulated based on our TB calculations; in addition to the experimentally determined hopping parameters, we include (Lorentzian) broadening of all the levels in the simulation (further details can be found in Appendix~\ref{TB}). The dispersive and flat bands can be clearly visualized in the spatially resolved d$I$/d$V$ maps. The low energy map shows the dispersive band, while the flat band is visible at the bias corresponding to the energy close to the on-site energy (in line with TB predictions). Note that the flat band lives exclusively on the rim sites, but the finite energy broadening means that we always sample some of the dispersive band as well, which explains the small intensity on the hub sites both in experimental and simulated LDOS. This effect is illustrated via further simulations in Appendix~\ref{TB}.

Similarly, the dispersive and flat bands of a cross lattice chain are visualized in the d$I$/d$V$ maps taken at the bias voltages of 3.15 V and 3.5 V, respectively, as shown in the Fig.~\ref{fig:2}(b). As discussed above, the NNN hoppings have an effect on the flat band wavefunction. If only NN hoppings are included, one would expect almost the same LDOS intensity on the site 1, 2, and 4. Including NNN hoppings gives zero intensity on sites 1 and 3 for the flat band. The finite energy resolution in our experiment means that we sample some of the dispersive band as well, which gives rise to slight LDOS intensity on site 1. This can be seen both in the experimental and simulated LDOS maps in Fig.~\ref{fig:2}(b) at a bias of 3.5 V. Comparison between the simulated response with and without NNN hoppings are shown in Appendix~\ref{NN-vs-NNN}. Our experimental results is in very good agreement with the simulated response including NNN interactions as can already be seen in Fig.~\ref{fig:2}(b). The difference between the diamond and cross chains is the gapped band structure of the cross chain and the presence of the gap can be clearly seen d$I$/d$V$ point spectra shown in Figs.~\ref{fig:2}(c,d) (simulated point spectra are shown in Appendix~\ref{point-spectra}). The spectra were acquired on the sites corresponding to the highest LDOS intensity on the dispersive (green) or the flat bands (purple) of the diamond and the cross lattices, respectively. In the case of diamond chain, the Dirac-like dispersive band has zero gap and correspondingly, the d$I$/d$V$ spectrum corresponding to the dispersive band is similar to the flat band spectrum due to the energy broadening (the spectrum measured at a site with the highest flat band LDOS has slightly narrower spectral width). On the other hand, the spectral features at the maximum dispersive band LDOS are clearly shifted to lower energies for the cross lattice reflecting the energy separation between the dispersive and the flat bands. 

\begin{figure}[t!]
    \centering
    \includegraphics{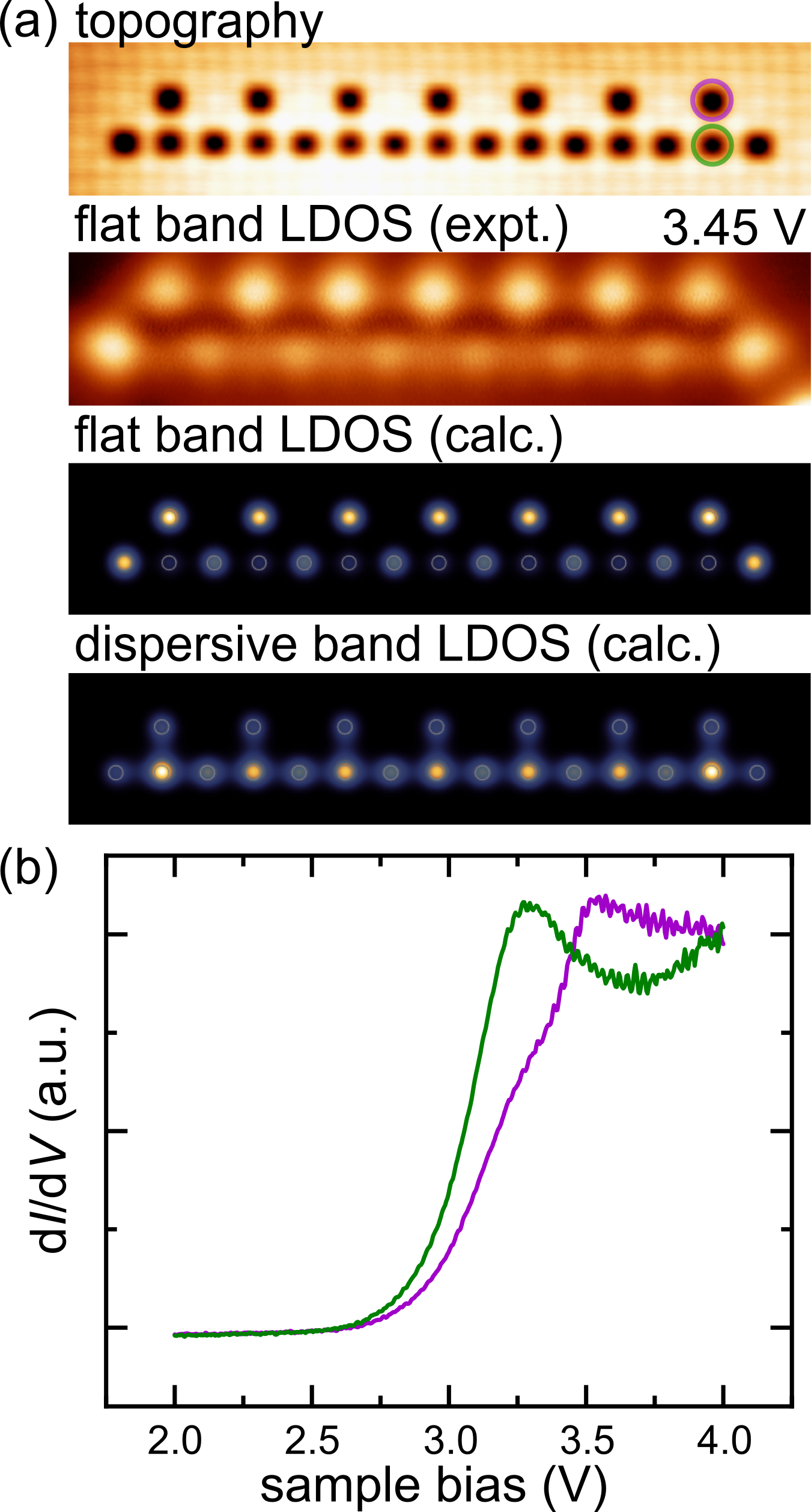}  
    \caption{Experimental realization of the 1D stub lattice. (a) Topography of the 1D stub lattice (top) and the experimental (measured close to the on-site energy) and simulated flat band LDOS maps (middle) along with the dispersive band LDOS simulation (bottom). (b) d$I$/d$V$ point spectra taken on flat band site (purple) and dispersive band site (green) as indicated in panel a.}
    \label{fig:3_5}
\end{figure}

\begin{figure}[t!]
    \centering
    \includegraphics{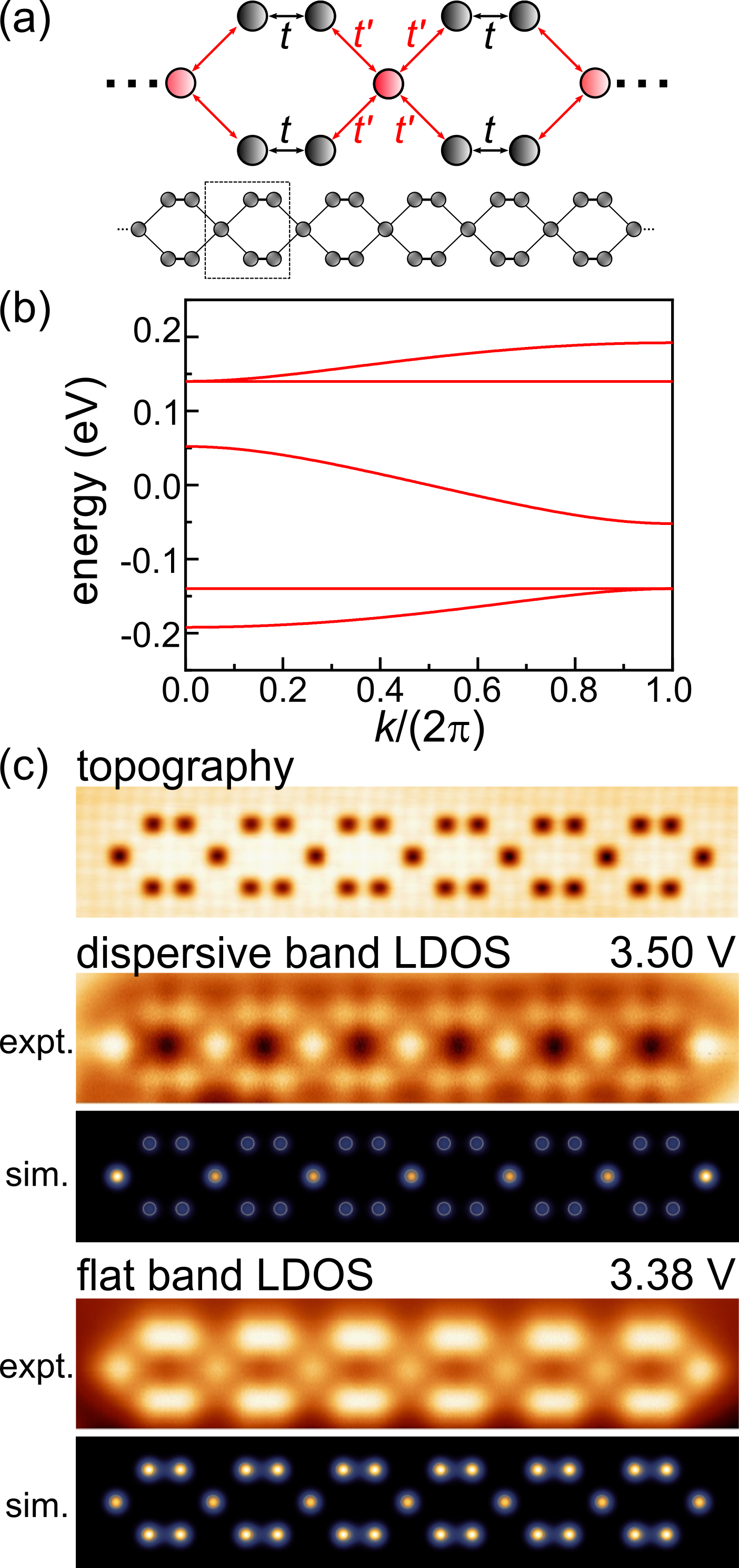}   
    \caption{1D extended diamond chain with multiple flat bands. (a) Schematic of lattice structure with a zoom-in on the connector site. (b) The calculated band structure of extended diamond chain (energies w.r.t.~the on-site energy). (c) Dispersive and flat band LDOS maps at the indicated bias voltages. The topography (top) and the calculated LDOS maps are given for reference.}
    \label{fig:3}
\end{figure}

In addition to the cross and diamond flat band lattices, we have also fabricated 1D stub lattice and characterised its electronic structure using d$I$/d$V$ spectroscopy and mapping. Fig.~\ref{fig:3_5} shows the experimental realization of the 1D stub lattice. Along with the topography, the Fig.~\ref{fig:3_5}(a) shows the experimental flat band LDOS measured by d$I$/d$V$ map taken at bias 3.45 V in the constant height mode which is in-line with simulated LDOS map. The calculated dispersive band LDOS map is also shown in the same figure (bottom). The d$I$/d$V$ point spectra is acquired on the site corresponding to flat band (purple) and dispersive band sites (green) show a clear energy gap between these two bands (see Fig.~\ref{fig:3_5}(b)).

\begin{figure*}[t!]
    \centering
    \includegraphics{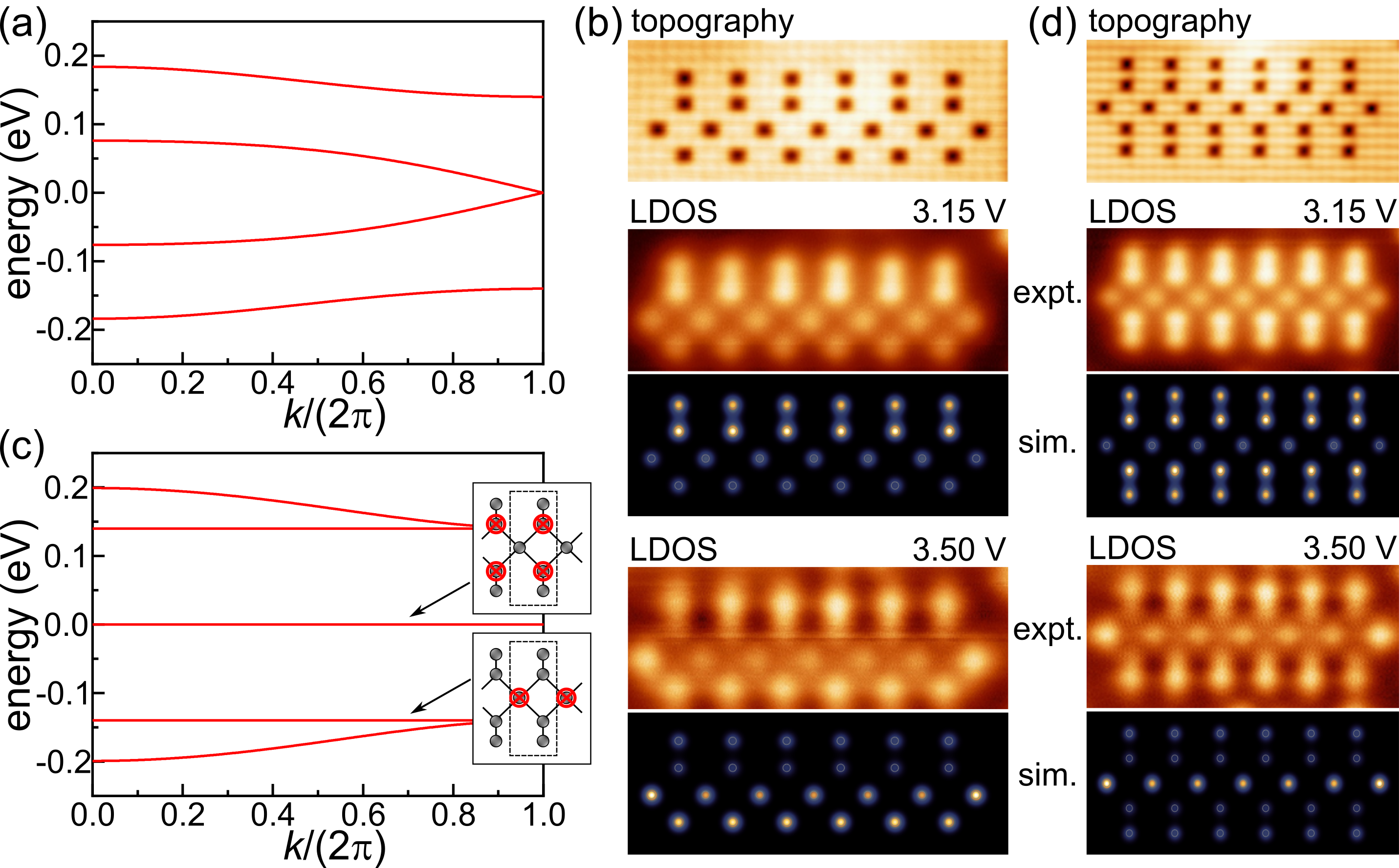}    
    \caption{Experimental control of the presence of the flat bands. (a) Calculated band structure of an asymmetric modified diamond chain. (b) Dispersive and flat band LDOS maps of the asymmetric modified diamond chain at the indicated bias voltages. The topography and simulated LDOS maps at corresponding energies are given for reference. (c) Calculated band structure of the symmetric modified diamond chain. The insets indicate the sites with zero wavefunction amplitude for the different flat bands. (d) Dispersive and flat band LDOS maps of the symmetric modified diamond chain at the indicated bias voltages. Energies in the calculated band structures are given w.r.t.~the on-site energy. The topography and simulated LDOS maps at corresponding energies are given for reference. }
    \label{fig:4}
\end{figure*}

\subsection{Multiple flat bands in an extended diamond chain}

As we discussed earlier, the flat bands may not survive when NNN hoppings are taken into account. For example, this causes the cross chain to only have one flat band. The number of dispersionless bands of a given lattice depends on the unit cell and specifically on the arrangement of the lattice sites around the connector sites \cite{Morales-inostroza2016}. Therefore, by analyzing the specific geometry, it is possible to construct different lattices having more than one flat band in the presence of NNN interactions. For example, we can consider the chain geometry shown in Fig.~\ref{fig:3}(a) forming an extended diamond lattice structure. The calculated energy spectrum of this lattice is shown in the Fig.~\ref{fig:3}(b). This extended diamond chain, unlike the diamond and cross lattice, has two flat bands at the energy $\pm$0.14 eV (for the hopping values corresponding the experimental system, in general the flat band position is $\pm t$ away from the on-site energy). Fig.~\ref{fig:3}(c) shows the experimental realization of this extended diamond chain. As shown in the Fig.~\ref{fig:3}(c) (middle), the d$I$/d$V$ LDOS map acquired close to on-site energy (3.50 V) depicts that the middle dispersive band is localized on the connector sites. On the other hand, the flat bands are localized on the miniarrays sites as shown in the d$I$/d$V$ LDOS map taken at 3.38 V energy (see Fig.~\ref{fig:3}(c) (down)). The flat band has zero wavefunction intensity on the connector sites, but again, due to the energy broadening in the experimental system, we pick some intensity from the (top and bottom) dispersive bands that have wavefunction intensity on all lattice sites.

\subsection{Controlling flat bands in symmetric and asymmetric modified diamond chains}
As the possible presence of flat bands depends on the geometry of the lattice \cite{deng2003}, it is possible to turn on/off the flat bands by maintaining/breaking the mirror symmetry of chain. For example, consider the diamond lattice (see Fig.~\ref{fig:2}(a)) having a flat band at the on-site energy. Now, if we break the symmetry by adding extra sites on one side of the diamond chain as shown in the Fig.~\ref{fig:4}(a,b), the calculated energy spectrum (see the Fig.~\ref{fig:4}(a)) shows that flat band splits into two dispersive bands. Fig.~\ref{fig:4}(b) shows the d$I$/d$V$ LDOS map of dispersive bands acquired below and at the on-site energy, respectively. Again, it is possible to regain the symmetry by adding the same number of extra sites on the other side of the asymmetric chain which results in three completely flat bands along with the two dispersive bands as shown in the Fig.~\ref{fig:4}(c). The experimental d$I$/d$V$ LDOS maps acquired at bias voltages of 3.14 V and 3.5 V shown in the Fig.~\ref{fig:4}(d) show that the lower energy flat band is localized on the added sites, while the flat band at the on-site energy sits on the center row of the chain, which is in-line with the simulated LDOS maps. The flat bands formed in this modified diamond chain are formed by having zero wavefunction amplitude at different connecting lattice sites. This is illustrated in the inset of Fig.~\ref{fig:4}(c), where the red symbols indicate which sites have zero wavefunction intensity. This is directly reflected in the experimental d$I$/d$V$ maps shown in Fig.~\ref{fig:4}(d). The simulated LDOS maps are in excellent agreement for the lower band for both the symmetric and asymmetric modified diamond chains (middle panels in Figs.~\ref{fig:4}(b,d)). On the other hand, the outer-most sites have some extra intensity in the experimental LDOS maps compared to the simulated maps for the middle flat band (bottom panels in Figs.~\ref{fig:4}(b,d)). This could be due to some additional energy broadening; nevertheless, the order of the relative intensities of the various sites compares nicely with the simulated LDOS maps.

\section{Conclusions}
We have successfully demonstrated the construction of atomically precise 1D chains exhibiting flat electronic bands. Depending on the exact geometry, the next-nearest neighbour interactions can cause flat-bands to become dispersive. We test the effects of NNN interactions and show which geometries show robust flat bands. Further modifications to the unit cell allow tuning the number and energy position of the flat bands and finally, they can be turned ``on'' and ``off'' by breaking and restoring the symmetry of the chain geometry. Our results constitute first steps in designing flat-bands in 1D solid-state systems and open the way towards detailed study of e.g.~topological flat band systems and experimental tests of flat-band-enhanced symmetry-breaking phase transitions.

\acknowledgements
This research made use of the Aalto Nanomicroscopy Center (Aalto NMC) facilities. We acknowledge support from the European Research Council (ERC-2017-AdG no.~788185 ``Artificial Designer Materials''), Academy of Finland (Academy professor funding no.~318995 and 320555, and Academy postdoctoral researcher no.~309975), and the Aalto University Centre for Quantum Engineering (Aalto CQE). 

\appendix

\section{Experimental methods}
\label{Expt}
	
All sample preparations and experiments were carried out in an ultrahigh vacuum system with a base pressure of $\sim$10$^{-10}$ mbar. The (100)-terminated copper single crystal was cleaned by repeated cycles of Ne$^{+}$ sputtering at 1.5\,kV, annealing to 600\,$^{\circ}$C. To prepare the chloride structure, anhydrous CuCl$_2$ was deposited from an effusion cell held at 300$^{\circ}$C onto the warm crystal ($T\approx$ 150 - 200$^{\circ}$C) for 180 seconds. The sample was held at the same temperature for 10 minutes following the deposition.
	
After the preparation, the sample was inserted into the low-temperature STM (Unisoku USM-1300) and all subsequent experiments were performed at $T=4.2$ K. STM images were taken in the constant current mode. d$I$/d$V$ spectra were recorded by standard lock-in detection while sweeping the sample bias in an open feedback loop configuration, with a peak-to-peak bias modulation of 20~mV at a frequency of 709~Hz. Line spectra were acquired in constant height; the feedback loop was not closed at any point between the acquisition of the first and last spectra. Manipulation of the chlorine vacancies was carried out as described previously \cite{Kalff:NatNano2016,Drost:NatPhys2017,Huda2020_npj}. The tip was placed above a Cl atom adjacent to a vacancy site at 0.5~V bias voltage and the current was increased to 1 to 2~$\mu$A with the feedback circuit engaged. The tip was then dragged towards the vacancy site at a speed of up to 250~pm/s until a sharp jump in the $z$-position of the tip was observed. This procedure lead to the Cl atom and the vacancy site exchanging positions with high fidelity.

\section{Tight-binding calculations}\label{TB}

\begin{figure}[!t]
\centering
\includegraphics [width=\columnwidth] {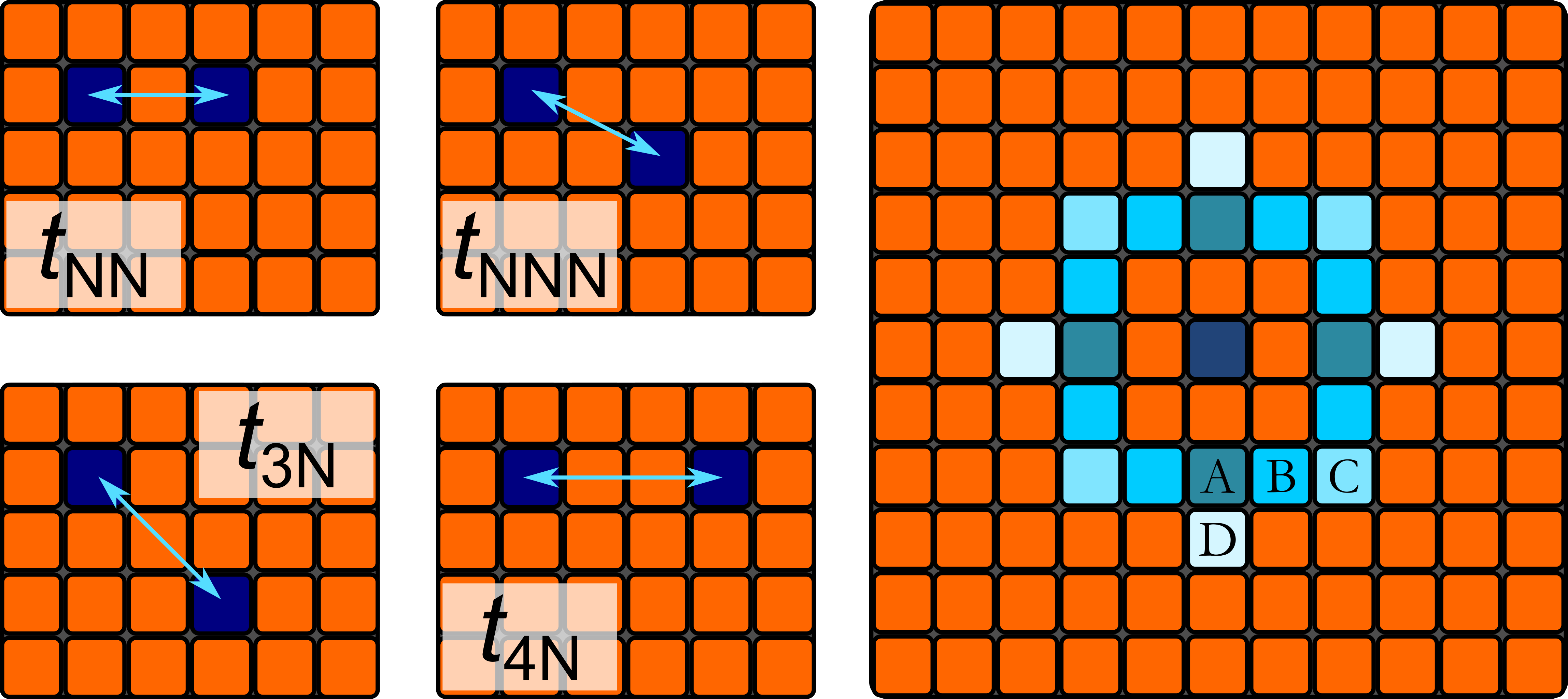}
\caption{Sketch of the c($2\times2$)-Cl adsorption structure and the different vacancy dimers. The individual panels show the different dimer configurations we have used. The larger schematic illustrates the hoppings between the central vacancy (dark blue) and the different symmetry-equivalent near-by vacancy sites with same color indicating a given value of the hopping. Letters A-D correspond to the different hoppings: $t_\mathrm{NN}=-0.14$ eV (A), $t_\mathrm{NNN}=-0.07$ eV (B), $t_\mathrm{3N}=-0.05$ eV (C), and $t_\mathrm{4N}=-0.04$ eV (D). }
\label{fig:SI_TB}
\end{figure} 

\begin{figure}[t!]
    \centering
    \includegraphics[width=\columnwidth]{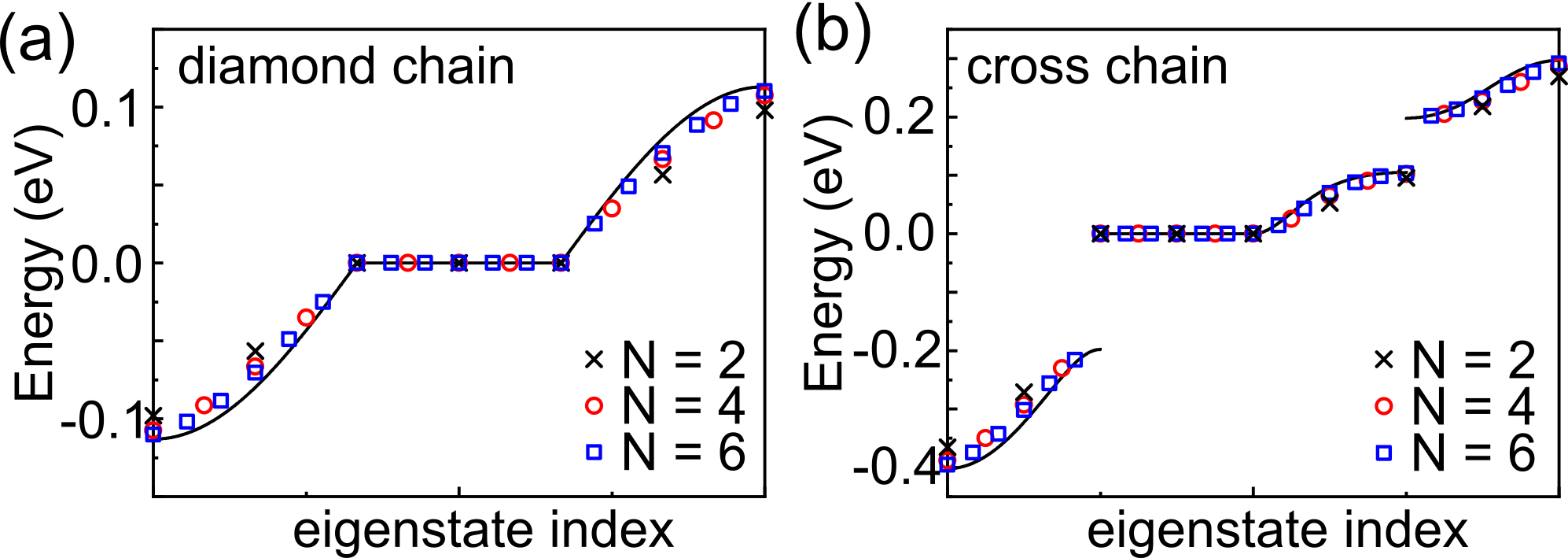} 
    \caption{(a,b) Effect of the system size ($N$ is the number of unit cells) on the energy levels of finite diamond (a) and cross chains (b). Solid lines give the result for a periodic system. The energy of the states belonging to the flat band is independent of system size. Energies are given w.r.t.~the on-site energy.}
    \label{fig:SI_size}
\end{figure}

The tight binding model is parametrised on the basis of our earlier work \cite{Drost:NatPhys2017}. The numerical values of the hopping amplitudes are (see also Fig.~\ref{fig:SI_TB}): $t_\mathrm{NN} = -0.14$ eV, $t_\mathrm{NNN} = -0.07$ eV, $t_\mathrm{3N} = -0.05$ eV, $t_\mathrm{4N} = -0.04$ eV. Atomic positions can be entered into a graphical user interface coded in Matlab representing the chlorine adsorption structure on Cu(100) to simulate structures of interest. The Matlab code implements the tight binding Hamiltonian 
\begin{equation}
    \mathscr{H}=-\sum_{ij} t_{ij} \hat{c}_i^{\dagger} \hat{c}_{j}+ h.c.
\end{equation}
where the summation runs over all pairs of sites and the hopping amplitudes are non-zero for the terms shown in Fig.~\ref{fig:SI_TB}. The on-site energy is set to $3.49\pm0.01$ V (value obtained experimentally from spectroscopy on a single vacancy). The resulting hopping matrix is diagonalised to obtain the eigenvectors and values. 

The effect of the finite size of the experimental structures is evaluated via TB calculations in Fig.~\ref{fig:SI_size}. It can be seen that already 6 unit cell long chains correspond rather well to the bulk energy dispersion.

Simulated local density of states (LDOS) maps at an energy $\epsilon_0$ are drawn according to:
\begin{multline}
    \rho(x,y,\epsilon_0) = \sum_j \left[ \sum_{i = 1}^{N} v_i(\epsilon_j) \mathrm{e}^{-((x-x_i)^2 - (y-y_i)^2)/(2\Gamma^2)} \right]^2 \\ \times \frac{1}{\pi}\frac{\Delta}{(\epsilon_j-\epsilon_0)^2+\Delta^2}
\end{multline}
where $j$ is the eigenstate index, $N$ is the total number of vacancy sites in the system,  $\epsilon_j$ and $v(\epsilon_j)$ are the energy and eigenvector corresponding to the $j^{th}$ eigenvalue, $(x_i,\ y_i)$ the position of the $i^{th}$ vacancy, $\Gamma$ is a phenomenological spatial broadening, and $\Delta$ the energy broadening.

\begin{figure}[t!]
    \centering
    \includegraphics{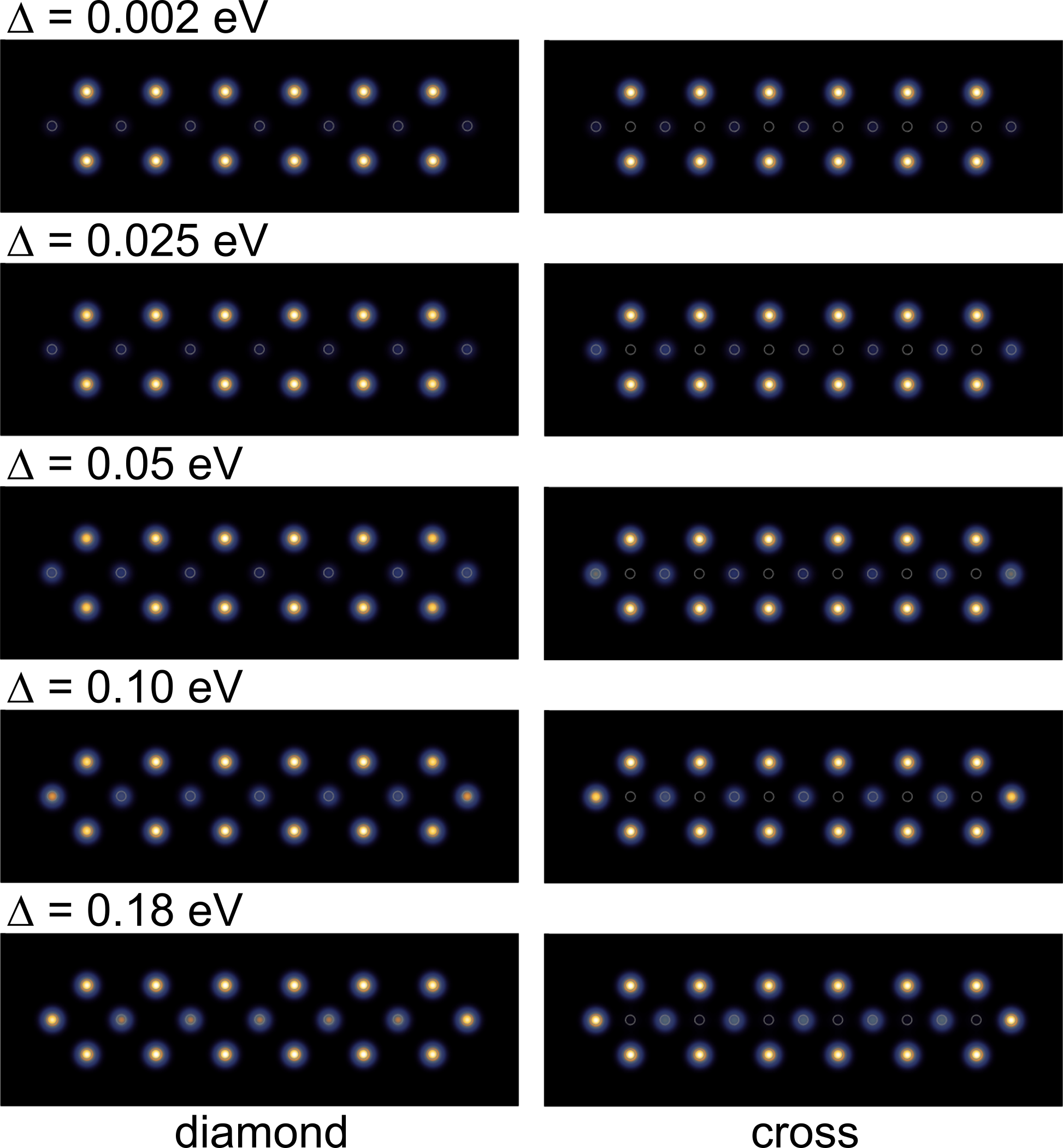} 
    \caption{Effect of the energy broadening on the simulated LDOS maps at the flat band energy (on-site energy) of the diamond (top) and cross (bottom) chains. Broadening causes the LDOS maps to have a contribution from the dispersive band.}
    \label{fig:SI_broadening}
\end{figure}

To obtain a simulated LDOS contour, for each eigenvalue, a two-dimensional Gaussian contour is placed at each vacancy site and scaled by the eigenvector entry corresponding to that site at a given eigenvalue. The complete map for that eigenvalue is squared and weighted with a lorentzian  according to difference between its eigenenergy and the energy of interest. The complete map is then the sum over all these constituents. The effect of the energy broadening is illustrated in Fig.~\ref{fig:SI_broadening} in the case of diamond and cross lattices.

Good agreement with the experimental data is reached for $\Delta = 0.18\pm0.01$ eV and $\Gamma = 0.71\ a$, where $a$ is the lattice constant of the c($2\times2$)-Cl structure. 

\begin{figure}[t!]
    \centering
    \includegraphics{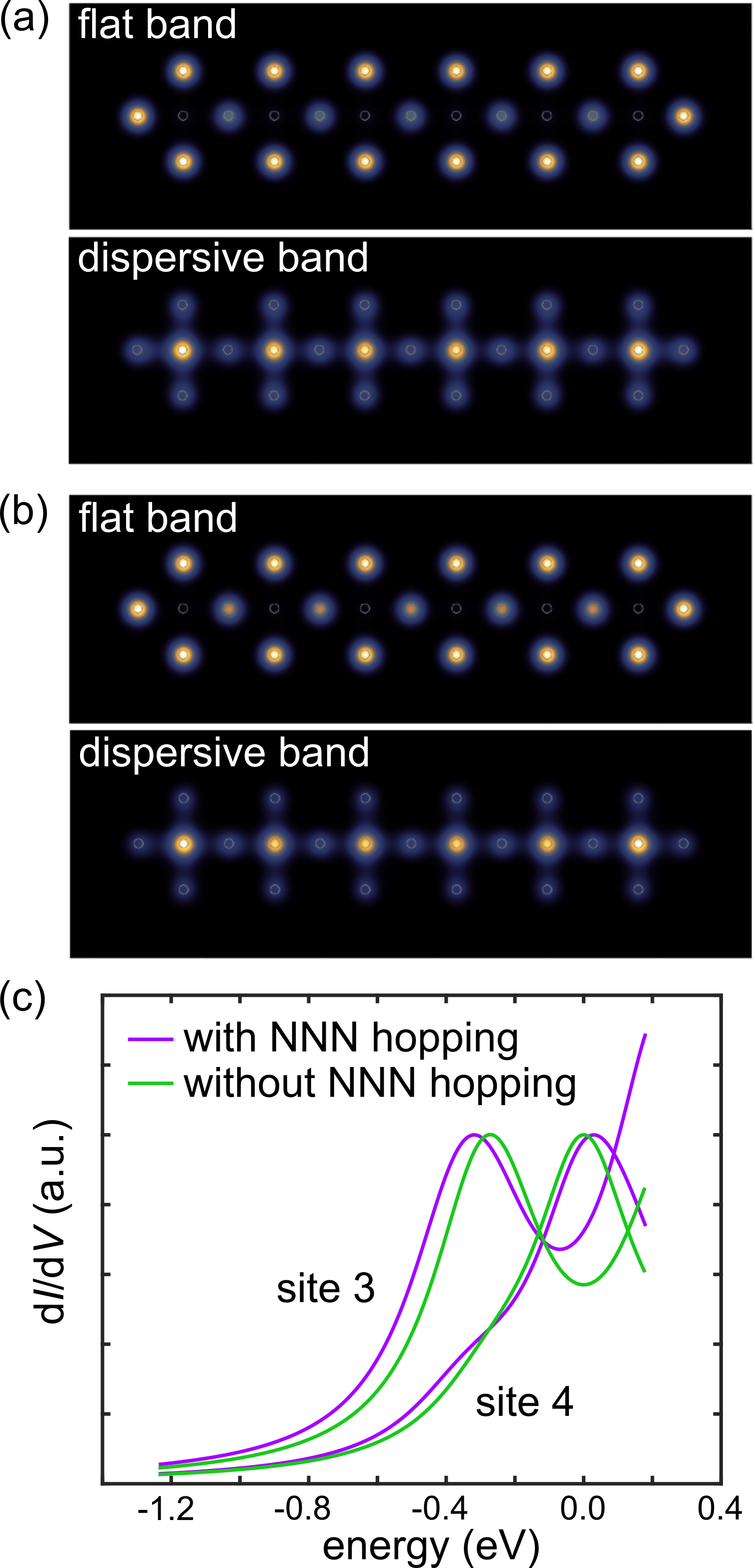}  
    \caption{Simulation of the LDOS maps and spectra of cross lattice. (a,b) Simulated LDOS maps of the flat and dispersive bands of the cross lattice with (a) and without (b) NNN hoppings. (c) Simulated d$I$/d$V$ spectra taken on the flat (site 4) and dispersive band sites (site 3) with and without NNN hoppings in the cross lattice structure. All the LDOS maps and spectra are simulated with the energy broadening corresponding to the experimental results ($\Delta=0.18$ eV). Energies are given w.r.t.~the on-site energy.}
    \label{fig:SI_NNN}
\end{figure}

\begin{figure}[t!]
    \centering
    \includegraphics{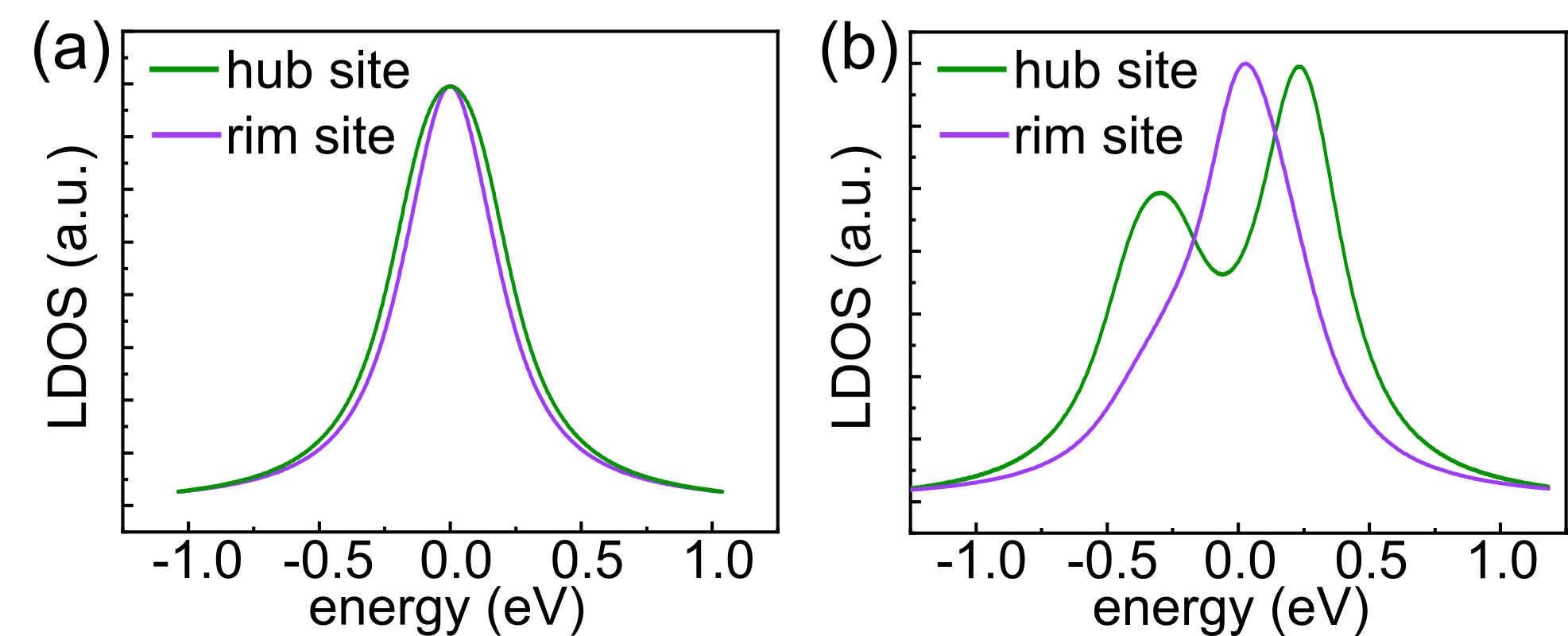}  
    \caption{(a,b) Simulated LDOS based on the TB model on the sites corresponding to the experimental spectra. The energy axis is w.r.t.~the on-site energy (which is indicated by the dotted lines in the experimental spectra in panels c and d).}
    \label{fig:SI_point}
\end{figure}

\section{Effect of the NNN hoppings in a cross lattice}\label{NN-vs-NNN}

In the case of the cross lattice, we cannot avoid NNN hoppings in our experimental system. Considering only the nearest-neighbor hoppings, the flat-band has zero intensity on the hub sites (site 3 in Fig.~\ref{fig:1}(a)). If we include NNN interactions, then also the bridge site 1 has to have zero intensity to cause electron localization. In order to compare these observations with the experimental results, we simulate LDOS maps with finite energy broadening  ($\Delta=0.18$ eV) at the energies of the flat and dispersive bands with and without NNN hoppings (Fig.~\ref{fig:SI_NNN}(a,b)). In both cases, the flat band has highest LDOS intensity on the rim sites (sites 2 and 4, see Fig.~\ref{fig:1}(a)), consistent with the experiments (Fig.~\ref{fig:2}(b)). With the NNN hoppings, the simulated LDOS maps have smaller intensity on the bridge sites 3 compared to the simulations without NNN hoppings. The experimental intensities are consistent with the simulated LDOS including the NNN interactions. The dispersive band LDOS is relatively insensitive to the NNN hoppings and has the highest intensity on the hub sites (site 3). Including NNN hoppings increases the simulated LDOS on the rim sites (2 and 4). The simulated d$I$/d$V$ spectra reflect these changes as well, see Fig.~\ref{fig:SI_NNN}(c).

\section{Simulated spectra on diamond and cross chains}\label{point-spectra}

Fig.~\ref{fig:SI_point} shows simulated spectra on the diamond and cross chains.

\clearpage
\newpage

\bibliography{bibliography}

\end{document}